%% file: main.tex
\documentclass{article}

\usepackage{arxiv}

\usepackage[utf8]{inputenc} % allow utf-8 input
\usepackage[T1]{fontenc}    % use 8-bit T1 fonts
\usepackage{hyperref}       % hyperlinks
\usepackage{url}            % simple URL typesetting
\usepackage{booktabs}       % professional-quality tables
\usepackage{amsfonts}       % blackboard math symbols
\usepackage{nicefrac}       % compact symbols for 1/2, etc.
\usepackage{microtype}      % microtypography
\usepackage{lipsum}
\usepackage{graphicx, subcaption}
\usepackage{multirow}
\usepackage{multicol}
\usepackage{xcolor}
\graphicspath{{images/}}

\usepackage{subfiles}

\title{Choosing a sampling frequency for ECG QRS detection using convolutional networks}

\author{
  Ahsan Habib \\
  School of Information Technology\\
  Deakin University\\
  Geelong, Australia 3225 \\
  \texttt{mahabib@deakin.edu.au} \\
   \And
 Chandan Karmakar \\
  School of Information Technology\\
  Deakin University\\
  Geelong, Australia 3225 \\
  \texttt{karmakar@deakin.edu.au} \\
   \And
 John Yearwood \\
  School of Information Technology\\
  Deakin University\\
  Geelong, Australia 3225 \\
  \texttt{john.yearwood@deakin.edu.au} \\
}

\begin{document}
\maketitle

\begin{abstract}
\subfile{sections/abstract}
\end{abstract}

% keywords can be removed

\section{Introduction}
\subfile{sections/introduction}

\section{Methodology} \label{method}

\subfile{sections/methodology}

\section{ECG Data} \label{ecgdata}

\subfile{sections/ecgdata}

\section{Results} \label{results}

\subfile{sections/results}

\section{Discussion} \label{discussion}

\subfile{sections/discussion}

\section{Conclusion and Future Work} \label{conclusion}
\subfile{sections/conclusion}

\bibliographystyle{unsrt}
\bibliography{main}

\end{document}

%% file: sections/abstract.tex
Automated QRS detection methods depend on the ECG data which is sampled at a certain frequency, irrespective of filter-based traditional methods or convolutional network (CNN) based deep learning methods.
These methods require a selection of the sampling frequency at which they operate in the very first place.
While working with data from two different datasets, which are sampled at different frequencies, often, data from both the datasets may need to resample at a common target frequency, which may be the frequency of either of the datasets, or could be a different one.
However, choosing data sampled at a certain frequency may have an impact on the model's generalisation capacity, and complexity.
There exists some studies that investigate the effects of ECG sample frequencies on traditional filter-based methods, however, an extensive study of the effect of ECG sample frequency on deep learning-based models (convolutional networks), exploring their generalisability and complexity is yet to be explored.
This experimental research investigates the impact of six different sample frequencies (50, 100, 250, 500, 1000, and 2000Hz) on four different convolutional network based models' generalisability (subject-wise intra and inter-database validation with three different datasets from PhysioNet bank) and complexity in order to form a basis to decide on an appropriate sample frequency for the QRS detection task for a particular performance requirement.
Intra-database tests report an accuracy improvement no more than approximately 0.6\% from 100Hz to 250Hz and the shorter interquartile range for those two frequencies for all CNN-based models.
The findings reveal that convolutional network-based deep learning models are capable of scoring higher levels of detection accuracies on ECG signals sampled at frequencies as low as 100Hz or 250Hz while maintaining lower model complexity (number of trainable parameters and training time).

%% file: sections/introduction.tex
Electrocardiogram (ECG) is a non-invasive technique of diagnosing heart conditions.
ECG signals consist of a P wave, a QRS complex and a T wave, among which the QRS complex is the most prominent and can easily be detected by a cardiologist.
The automated analysis of ECG signals requires these basic ECG waves to be detected and the detection of the QRS complex is often the basis of detection of the other waves \cite{Kohler2002}.
Sampling frequency of an ECG signal may characterise the quality of the signal in terms of the visibility of the basic waves.
A poorly chosen sampling frequency may have an impact on the ECG signal quality, as well as, on the performance of the automated ECG signal analysis, specifically, the QRS detection.

\begin{figure}[h]
  \centering
  \begin{subfigure}[b]{0.3\textwidth}
    \includegraphics[width=\textwidth]{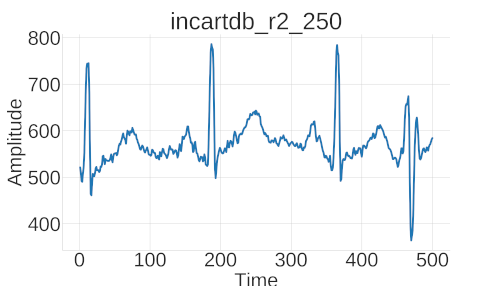}
    \caption{}
  \end{subfigure}
  \begin{subfigure}[b]{0.3\textwidth}
    \includegraphics[width=\textwidth]{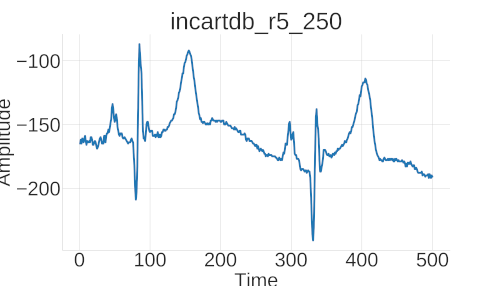}
    \caption{}
  \end{subfigure}
  \begin{subfigure}[b]{0.3\textwidth}
    \includegraphics[width=\textwidth]{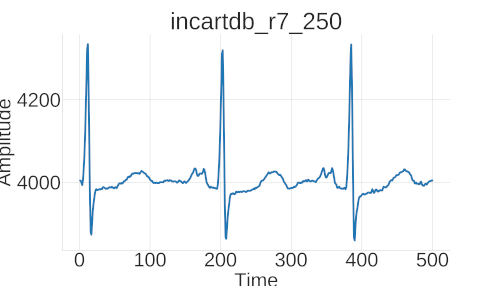}
    \caption{}
  \end{subfigure}
  \caption{Two seconds ECG signal excerpt of INCART dataset (resampled at 250Hz) of three patients with symptoms of (a) coronary artery disease, arterial hypertension (record number 2), (b) acute MI (record number 5), and (c) transient ischemic attack (record number 7).}
  \label{fig:ecg_data_incart}
\end{figure}

The ECG is a non-stationary (varying mean and variance), higly irregular, and heterogeneous time-series data~\cite{ganapathy_deep_2018}.
ECG signals vary continuously over time and reflects the clinical state of the human body.
Recording of the ECG signals generally contains different types of noise (i.e. baseline noise, power-line interference, electrode contact noise) and artifacts (i.e. motion artifacts) \cite{Elgendi2014}.
Also, different ECG recordings of the same subject in different recording environments, may produce recordings with different signal characteristics.
In addition, ECG recordings of each individual may be deemed to be so unique that it can be used as a bio-metric signature to indentify individuals \cite{Zhang2017}.
Due to the non-stationary nature of ECG signals, the inherent subject-wise signal variance and the presence of noise and motion artifacts, the publicly available ECG datasets (i.e. in the PhysioNet data bank) show diverse signal characteristics.
For example, the MIT-BIH arrhythmia database \cite{goldberger_physiobank_2000} consists of a mix of healthy and arrhythmia (with rare but clinically important types) patients, whereas, the INCART database \cite{goldberger_physiobank_2000} consists of recordings from patients consistent with ischemia, coronary artery disease, conduction abnormalities and arrhythmias (Figure \ref{fig:ecg_data_incart} shows three patients with different symptoms), and the QT database \cite{laguna_database_1997}, on the otherhand, includes ECGs representing a wide variety of QRS and ST-T morphologies and this is a compositional database (combining recordings from other PhysioNet databases). In this study, we aim to analyze the impact of sampling frequency on the performance of QRS detection algorithm.

\begin{figure}[h]
  \centering
  \begin{subfigure}[b]{0.3\textwidth}
    \includegraphics[width=\textwidth]{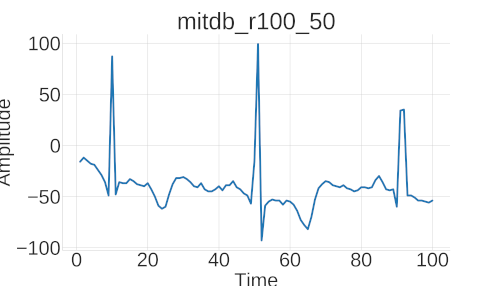}
    \caption{}
  \end{subfigure}
  \begin{subfigure}[b]{0.3\textwidth}
    \includegraphics[width=\textwidth]{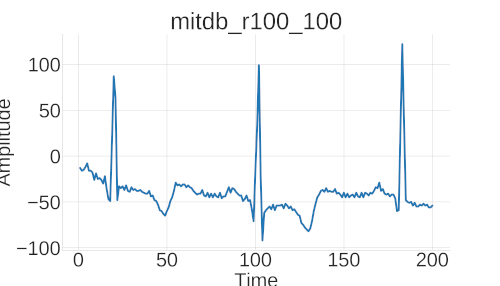}
    \caption{}
  \end{subfigure}
  \begin{subfigure}[b]{0.3\textwidth}
    \includegraphics[width=\textwidth]{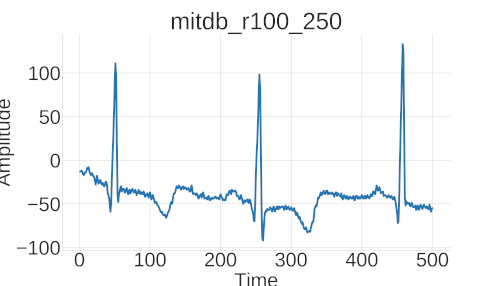}
    \caption{}
  \end{subfigure}
  \begin{subfigure}[b]{0.3\textwidth}
    \includegraphics[width=\textwidth]{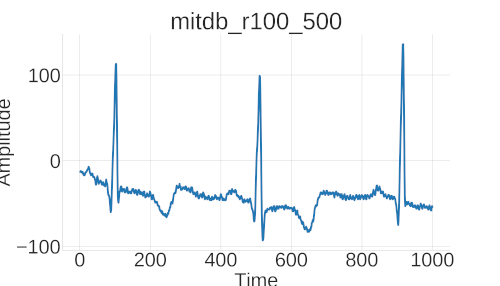}
    \caption{}
  \end{subfigure}
  \begin{subfigure}[b]{0.3\textwidth}
    \includegraphics[width=\textwidth]{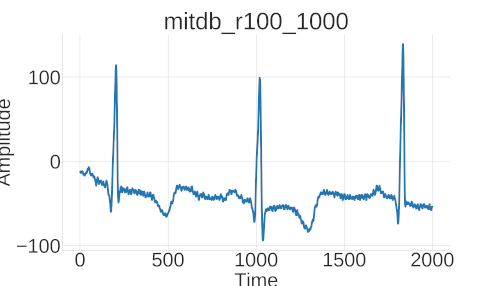}
    \caption{}
  \end{subfigure}
  \begin{subfigure}[b]{0.3\textwidth}
    \includegraphics[width=\textwidth]{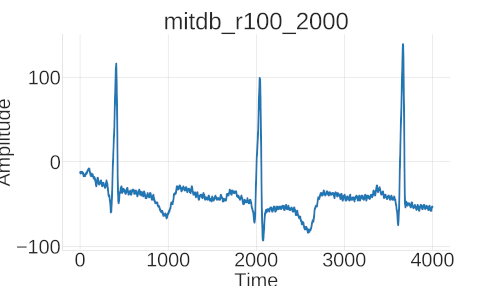}
    \caption{}
  \end{subfigure}

  \caption{Two seconds ECG signal excerpt of MIT-BIH arrhythmia dataset (record number 100), resampled at all six sample frequencies - (a) 50Hz, (b) 100Hz, (c) 250Hz, (d) 500Hz, (e) 1000Hz, and (f) 2000Hz. The original sample frequency is 360Hz, which was resampled using linear interpolation (using PhysioNet \emph{xform} tool). ECG data at the lowest frequency is less noisy but loses information, on the other hand, higher frequency signals are of finer resolution but comparatively more noisy due to interpolation.}
  \label{fig:ecg_data_viz}
\end{figure}

The impact of sampling frequency on the morphology of ECG signal (healthy individual) is shown in Figure \ref{fig:ecg_data_viz}.
This effect on the pathological or very noisy ECG signals may be worse, which can negatively affect the performance of any QRS detection algorithm.
The shape of the QRS complexes may be distorted at lower frequencies and the high frequency signals, which provide finer resolution, however, are likely to contain noise due to the subtle movements.
Thus, lower frequency signals may risk losing information that keeps an automated QRS detector from obtaining a desirable performance, and on the otherhand, much higher frequency signals may risk not gaining much information to increase detection performance proportionately and increase the algorithm complexity.
The selection of an optimal sample frequency for the QRS detection task is thus desirable to obtain a general detector algorithm.

The selection strategy of the target sample frequency is different in different studies, and includes, choosing the database frequency as the target in a single database-based study \cite{DeChazal2004a}, choosing one of the database frequeuncies as the target \cite{habib_impact_2019}, or choosing an arbitrary target frequency \cite{guo_inter-patient_2019}.
Automated ECG analysis methods, in the literature, often report their performances against a single target sample frequency.
Thus, choosing a single best or optimum target frequency, for a specific task, is an important design decision which may require having a clear understanding of the effects of the sampling frequency on the model's performance.

%To capture complex patterns (i.e., QRS complex detection), ECG signals sampled at higher frequencies may be desirable, but it is found that different ECG databases are sampled at different frequencies (i.e. BIH arrhythmia, INCART, and QT databases sampled at 360Hz, 257Hz, and 250Hz respectively).
%In addition, the selection strategy of the target sample frequency is different in different studies, which includes, to choose the database frequency as the target in the single database-based study \cite{DeChazal2004a}, to choose one of the database frequeuncies as the target \cite{habib_impact_2019}, or to choose an arbitrary target frequency \cite{guo_inter-patient_2019}.
%Automated ECG analysis methods, in the literature, often reports their performances against a single target sample frequency.
%Thus, choosing a single best or optimum target frequency, for a specific task, is an important design decision which may require to have a clear understanding of the effects of the sampling frequency on the model's performance.

There are different models (or approaches) for the QRS detection task, including, template matching \cite{nakai_noise_2014}, peak-detection-based methods using filters \cite{hamilton_quantitative_1986}, and deep learning (DL) based studies \cite{habib_impact_2019, chandra_robust_2019}.
Different methods' detection performance may vary based on the variation of the sample frequency due to their internal working mechanisms.
Filter-based traditional QRS detection methods, for example, use frequency dependent filters (i.e. band-pass filter) which are sample frequency specific and different studies tried to find the sample frequency effects on the model's performance \cite{ajdaraga_analysis_2017, ziemssen_influence_2008, mahdiani_is_2015} and to minimise them \cite{ajdaraga_analysis_2017, hamilton_open_2002}.
Deep learning-based methods may also have effects on the variation of the sampling frequency, though not the same way as the filter-based methods mentioned above, which should be investigated.

Deep learning-based approaches for the QRS detection task have been a promising research direction, inspired by success in computer vision.
DL-based methods may not be dependent on the data sampled at different frequencies the same way as the filter-based methods would be, however, there are several aspects of the deep learning-based algorithms those may be dependent on the sample frequency of the input signal.
It may become intuitive, for the DL-based methods for QRS detection, that due to the loss of information (occuring at the lower sample frequencies), detection performance may go down and on the otherhand, due to the cessation of information gain (occured at the higher sample frequencies), the increase of the detection performance may be zero or marginal, but a quantitative analysis seems to be necessary to spot these frequency boundaries and the detection behavior over this range of frequencies.
%DL-based models, specifically, convolutional networks (CNNs), are able to be trained on the raw pixel data for optical character recognition and handwriting recognition tasks and successfully been deployed in the industry by the late 1990s \cite{lecun_deep_2015}.
%It was also found that CNN shows a faster convergence in the backpropagation training in the presence of carefully injected noise compared to a noiseless backpropagation training, validated on the MNIST image dataset \cite{audhkhasi_noise-enhanced_2016}.
Two features of CNNs: i) having the capability to handle raw input data; and ii) the robustness to noise, make them different from the filter-based methods in detecting QRS complex from ECG signal with varying sampling frequency \cite{lecun_deep_2015,audhkhasi_noise-enhanced_2016}.
As mentioned earlier, the higher sampling frequencies that enable finer resolution input but likely introduce noise, may limit the information gain and thus, may limit the detection performance, but CNNs, equipped with noise tolerance, may show a different performance behavior, which should be properly investigated for the QRS detection.

In our previous work \cite{habib_impact_2019}, the impact of the diverse ECG datasets was investigated on the generalisation capability of a CNN model for the QRS detection task.
% The CNN-based model's generalisability was investigated using the subject-wise intra-database validation technique where all the patients belong to the same database (i.e. the MIT-BIH arrhythmia database).
% In addition, the same model was trained and validated using the same intra-database technique on the two other open ECG databases (i.e. INCART and QT), and it was found that the same model generalises differently on different databases.
% To investigate the same model's generalisability further, a cross-database validation was performed where the model trained on a single database, was validated on the other two databases, and it was found that the single model's generalisability differs across other validation databases depending on the training database.
The CNN-based model's generalisability was investigated using subject-wise intra-database and cross-database validation which showed that a single model generalises differently on different databases.
Finding a general model for the QRS detection task is desirable, however, not many studies in the literature were found to investigate a single model's generalisability across multiple ECG databases.
While working on three databases, which were sampled at three different frequencies, in that study, the same question of choosing an optimum target sample frequency was encountered and the 360Hz was chosen considering the fact that the MIT-BIH arrhythmia database, which was one of the mostly studied databases, was sampled at that frequency and achieved superior performances in many different studies.
However, a thorough study focusing on the effects of the sampling frequency on CNN-based models was missing during our previous work to help us choose an optimum target frequency.
In that study, the generalisabilty of only the proposed CNN model was investigated across multiple databases, however, those findings were not validated across multiple CNN model architectures to affirm the generalisability of DL models.

% A QRS detection algorithm, irrespective of types of approaches (i.e. filter-based or deep learning-based), has to decide a target sample frequency to work with and the train/test datasets are resampled at that frequency.
% CNN model-based studies for different ECG analysis tasks, as found in the literature, used different target sampling frequencies for different cases.
% For example, for the multi-frequency dataset-based studies, the target sample frequency is often one of the data frequencies \cite{habib_impact_2019} (i.e. target 360Hz from 360Hz, 257Hz, and 250Hz data) or arbitrary \cite{guo_inter-patient_2019} (i.e. target 180Hz from 360Hz and 128Hz data), and on the other hand, for the single dataset-based studies, the target sample frequency is often the same as the data frequency \cite{DeChazal2004a} or is rather arbitrary.
% In the previous study, mentioned above, the target sample freuqency was 360Hz, at which one of the three datasets (i.e. the MIT-BIH arrhythmia dataset) was sampled.
% The analysis of the CNN model's generalisation capability for the QRS detection on different sample frequency data seems to be an obvious next step of the previous study.

The current study examines the impact of six different sampling frequencies, between as low as 50Hz and as high as 2000Hz (i.e. 50Hz, 100Hz, 250Hz, 500Hz, 1000Hz, and 2000Hz), on the generalisabilty of a CNN model.
% on three different aspects of the CNN based-models - (i) generalisability, (ii) selection of model hyperparameters (i.e. convolution filter size), and (iii) the computation complexity.
Three publicly available ECG datasets (sampled at three different frequencies), from the PhysioNet databank, were used (i.e. MIT-BIH arrhythmia, INCART, and QT) and subject-wise intra and inter-database validations were carried out for the models generalisability analysis.
While the intra-database testing reveals the model's generalisability for the subjects within a dataset, the inter-database test validates a model's generalisability beyond the training dataset.
% Each ECG dataset is unique in terms of the sample frequency, intra and inter-patient ECG data variability, and the recording environment.
% MIT-BIH arrhythmia database \cite{goldberger_physiobank_2000} consists of a mix of healthy and arrhythmia (with rare but clinically important types) patients.
% INCART database \cite{goldberger_physiobank_2000}, on the otherhand, consists of recordings from patients consistent with ischemia, coronary artery disease, conduction abnormalities and arrhythmias (Figure \ref{fig:ecg_data_incart} shows three patients with different symptoms).
% The QT database \cite{laguna_database_1997}, last but not the least, includes ECGs representing a wide variety of QRS and ST-T morphologies and this is a compositional database (combining recordings from MIT-BIH arrhythmia database, the European Society of Cardiology ST-T Database, and several other ECG databases collected at Boston's Beth Israel Deaconess Medical Center).
Selection of three datasets, in this study, was assumed to ensure the selection of ECG signals with a diverse range of cardiac (patho) physiological characteristics for the QRS detection task.
In order to cross validate the findings of the sampling frequency effects on DL models, four different CNN architectures were used (classical convolution-pooling, Fully convolutional networks (FCN) \cite{long_fully_2015, wang_time_2017}, Discriminative filter learning model (DFL) \cite{wang_learning_2018}, and DenseNet \cite{huang_densely_2017}), considering the assumption that these CNN architectures cover a broad range of DL models.
% In this study, the performance variation of these CNN based-models across diversified sample frequencies is investigated.

%% file: sections/methodology.tex
\subsection{Data resampling}
The heart beats at most 220 beats per minutes (which is less than 4Hz), but there are heartbeat signals whose spectrum is up to 25Hz or beyond \cite{ajdaraga_analysis_2017}.
For example, a small QRS complex may be 40 ms that corresponds to 25Hz.
The Nyquest theorem dictates that the input signal can represent up to half of its sampling frequency \cite{proakis_digital_2004}.
This essentially means that the sampling frequency of the ECG signal should be at least 50Hz to represent a small QRS complex spectrum of 25Hz.
To analyse the effect of different sampling frequencies on the CNN model generalisation, six different target sample frequencies were chosen from as low as 50Hz to as high as 2000Hz (i.e., 50, 100, 250, 360, 500, 1000 and 2000Hz).
ECG signals below 50Hz may lose information and make the signal unsuitable for the QRS detection task, whereas, a 2000Hz signal represents enough finer resolution, beyond which no additional information may be added to the signal for the specified task.
The PhysioNet provided function (\emph{xform}) was found suitable to resample all three datasets (MIT-BIH arrhythmia, INCART, and QT) to six target frequencies, with scaling the annotation data accordingly.

\subsection{Data segmentation}
To localise heartbeats, the ANSI/AAMI EC38 and EC57 standards \cite{ecar_aami_recommended_1987} suggest that an estimated location should be considered accurate if it is within 150 ms from the corresponding annotated location.
This implies that a location can be considered as an R-peak if the corresponding annotations point stays around 150 ms of that point, forming a valid window range of 300 ms \cite{chandra_robust_2019}.
ECG recordings from the datasets were segmented with a length of 300 ms equivalent samples following this policy and an overlap of 50\% was considered to shift the segment forward.
Raw ECG segments were fed without any pre-processing to the CNN model for QRS detection.

\subsection{CNN Models}

To justify the impact of different sampling frequency signals on the CNN models, it was important to use more than one CNN architecture, to understand the general model behavior across different CNN model architectures.
Four different CNN model architectures were considered in this study, as below -
\begin{itemize}
  \item Classical convolution-pooling model (alternate conv-pooling)
  \item DenseNet
  \item Fully convolutional network (FCN)
  \item Discriminative filter learning model (DFL)
\end{itemize}
Note that there are a plathora of CNN model architectures available in the literature, and considering all of them is impractical, however, some successful model architectures from the computer vision domain may not be suitable for the current context where comparatively shallower CNN models are preferrable.
For example, an architecture applying residual connection (i.e. ResNet CNN architecture \cite{he_deep_2016}) may generally not be suitable for a shallow CNN model with less than eight layers of convolutions \cite{hannun_cardiologist-level_2019}, and thus, not considered for the current study.
To capture the general behavioral pattern of CNN models across different sample frequency data was the main goal in this study, and it was assumed that the above mentioned four CNN model architectures would represent a broader CNN model family.

\begin{figure}[h]
  \centering
  % \begin{subfigure}[b]{0.9\textwidth}
  %   \hspace{1in}
  %   \includegraphics[scale=0.3]{arch_four}
  % \end{subfigure}
  \begin{subfigure}[b]{0.35\textwidth}
    \includegraphics[width=\textwidth]{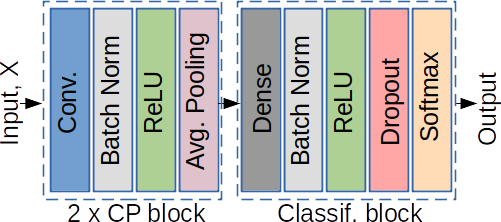}
    \caption{}
  \end{subfigure}
  \begin{subfigure}[b]{0.35\textwidth}
    \includegraphics[width=\textwidth]{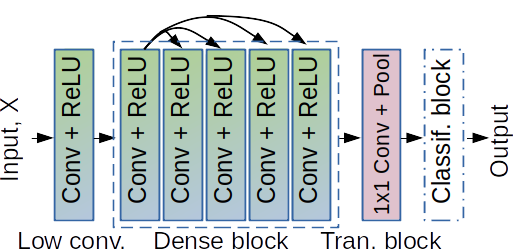}
    \caption{}
  \end{subfigure}
  \begin{subfigure}[b]{0.35\textwidth}
    \includegraphics[width=\textwidth]{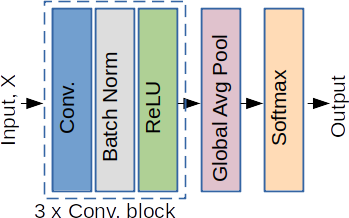}
    \caption{}
  \end{subfigure}
  \begin{subfigure}[b]{0.35\textwidth}
    \includegraphics[width=\textwidth]{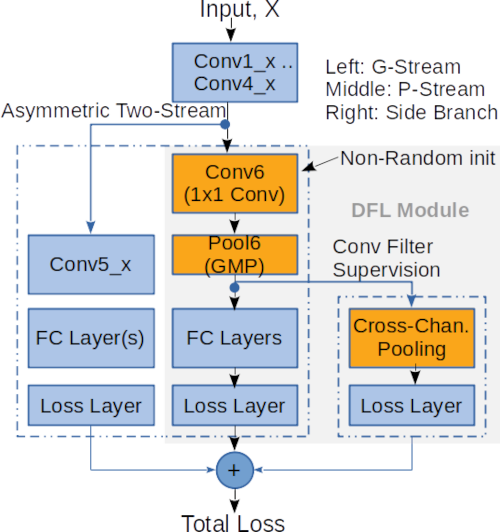}
    \caption{}
  \end{subfigure}
  \caption{Four CNN models used in this study, (a) Classical convolution-pooling network (b) densely connected network (DenseNet), (c) Fully Convolutional network, and (d) discriminative Filter-bank learning (DFL). Note that the Classical conv-pooling network (Figure \ref{fig:arch_four}-a) consists of two sets of convolution-pooling (CP) pair, followed by a classification block, which consists of a hidden dense layer and the final softmax layer.The DenseNet (Figure \ref{fig:arch_four}-b) consists of a single densely connected block (of five convolution layers with proper padding to maintain the input and output resolution the same) and a transition-block (bottle-neck layer, followed by a pooling layer), followed by the same classification block of the Classical convolution-pooling network (Figure \ref{fig:arch_four}-a). The Fully convolutional network (Figure \ref{fig:arch_four}-c) consists of three convolution layers (convolution + batch-normalisation + activation) with proper padding (to maintain the input and output size constant), followed by a global-average-pooling and softmax layer. The DFL \cite{wang_learning_2018} is an approach to enhance the mid-level representation learning within the CNN framework, by learning a bank of convolutional filters (Figure \ref{fig:arch_four}-d). This consists of a global-stream (G-stream at the left branch, five Convolution layers, followed by a Dense and loss layer), discriminative patch stream (P-stream, 1x1 convolution followed by a global-max-pooling, Dense, and loss layer), and finally, a righthand side branch of cross-channel pooling (followed by a loss layer) for convolutional filter supervison and finally, all these losses combine to produce the final loss.}
  \label{fig:arch_four}
\end{figure}

% \begin{figure}[h]
%   \centering
%   \begin{subfigure}[b]{0.4\textwidth}
%     \includegraphics[width=\textwidth]{arch_dfl}
%     % \caption{}
%   \end{subfigure}
%   \caption{The discriminative filter-bank learning (DFL) \cite{wang_learning_2018} is an approach to enhance the mid-level representation learning within the CNN framework, by learning a bank of convolutional filters. This consists of a global-stream (G-stream at the left branch, five Convolution layers, followed by a Dense and loss layer), discriminative patch stream (P-stream, 1x1 convolution followed by a global-max-pooling, Dense, and loss layer), and finally, a righthand side branch of cross-channel pooling (followed by a loss layer) for convolutional filter supervison. Finally, all these losses combine to produce the final loss.}
%   \label{fig:arch_dfl}
% \end{figure}

\subsubsection{Classical convolution-pooling network}
Classical convolution-pooling represents a CNN model where a pooling layer always follows a convolution layer, forming a pair and in this study, two such convolution-pooling pairs were used followed by a classification block, which consists of a hidden dense layer followed by an output layer of two neurons for the QRS or non-QRS decision (Figure \ref{fig:arch_four}-a).

\subsubsection{Dense network, DenseNet}
The Dense network model (DenseNet) is a densely connected convolutional network which connects each layer to every other layer in a feed-forward fashion \cite{huang_densely_2017} and consists of stacked dense blocks and transition layers.
Dense blocks may contain several densely connected convolution layers, whereas, the transition layer down-samples its input dimension by employing a bottle-neck and a pooling layer.
In this study a single dense block and a transition layer were used, where the dense block consisted of five densely connected convolution layers preserving the input resolution by applying proper padding (Figure \ref{fig:arch_four}-b).
Finally, a classification block was used, similar to that of the classical convolution-pooling architecture (Figure \ref{fig:arch_four}-a), to get the final output.

\subsubsection{Fully convolutional network, FCN}
The Fully convolutional network (FCN) was originally used for semantic segmentation of images \cite{long_fully_2015}, and later applied for time series classification \cite{wang_time_2017} where a sequence of convolution layers were used and fully-connected layers were replaced by a global average pooling (GAP) layer.
A similar approach was applied in this study where three convolutional blocks (i.e. a convolution block = convolution layer + batch normalisation + activation layer) were followed by a global-average-pooling (GAP) layer and finally a traditional softmax classifier was fully connected to the GAP layer's output (Figure \ref{fig:arch_four}c).

\subsubsection{Discriminative filter-bank learning, DFL}
A discriminative filter bank within a single convolutional network context can be trained for fine-grained recognition \cite{wang_learning_2018} in image data and found promising (Figure \ref{fig:arch_four}-d).
In this approach, the low-level convolution features are passed through two asymmetric streams - global feature learning stream (G-stream) and discriminative patch learning stream (P-stream) and additionally, forking a side branch from the global-max-pooling (GMP) output of the P-stream and passing through a cross-channel pooling for convolutional filter supervision.
As shown in the Figure \ref{fig:arch_four}-d, the 1x1 convolutional layer in the P-stream may not trigger at the discriminative patches, however, to do that a direct supervision was imposed at the 1x1 filters by introducing a cross-channel pooling layer followed by a softmax loss layer.
A one-dimensional counterpart of the approach was implemented for QRS detection task, in this study.

\subsection{Model hyper-parameters}
Model hyper-parameters generally require optimization to find the best working model for a particular dataset.
The focus of this study is to observe the effect of sampling frequency on the CNN model generalisation, instead of finding the best model.
Model hyper-parameters were chosen for a particular model keeping the number of trainable parameters as low as possible and at the same time, maintainiing higher accuracy levels.
Each of the four models went through such optimisations as described below and the selected models configurations are summarised in the Table \ref{tab:model_config}.

\begin{table}[h]
    \centering
    \caption{Configuration information of the four CNN models (Classical convolution-pooling, Fully convolutional network, Discriminative filter-bank learning, and DenseNet) used in this study.}
    \begin{tabular}{|c|l|c|}
        \hline
        \emph{Model} & \emph{Config name} & \emph{Config value} \\
        \hline
        \multirow{3}{*}{Classical CNN} & Conv-pooling pair & 2 \\ \cline{2-3}
        & Conv in, out-channel & 1, 5 \\ \cline{2-3}
        & Average pooling & /2 \\ \cline{2-3}
        & Hidden dense neurons & 50 \\ \cline{2-3}
        \hline
        \multirow{3}{*}{FCN} & Convolution blocks & 3 \\ \cline{2-3}
        & Conv channels per block & 128, 128, 2 \\ \cline{2-3}
        \hline
        \multirow{3}{*}{DFL} & No. of init conv layers & 4 \\ \cline{2-3}
        & Streams after low-conv & 2 (g/p-stream) \\ \cline{2-3}
        & Init. conv out-channel & 64 \\ \cline{2-3}
        & G-stream conv out-channel & 16 \\ \cline{2-3}
        & P-stream conv out-channel & 64 \\ \cline{2-3}
        \hline
        \multirow{3}{*}{DenseNet} & No. of dense blocks & 1 \\ \cline{2-3}
        & No. of transition layer & 1 \\ \cline{2-3}
        & No. of conv-layers in dense block & 5 \\ \cline{2-3}
        & Dense block conv out-channels & 32 \\ \cline{2-3}
        & Tran. layer conv out-channels & 32 \\ \cline{2-3}
        & Tran. layer avg pooling & /2 \\ \cline{2-3}
        & Hidden dense neurons & half of the flattened feat. \\ \cline{2-3}
        \hline
        * All models & Conv kernel size & 44 ms equiv. \\
        \hline
    \end{tabular}
    \label{tab:model_config}
\end{table}

\subsubsection{Classical convolution-pooling model parameters}
Classical CNN model generally consists of multiple pairs of alternate convolution-pooling layers where the use of a single convolution-pooling pair is the minimum.
Two pairs of convolution-pooling blocks were found better compared to the minimum configuration of a single pair-block and so, was used for this study.
The number of convolution output channels was set to five, since choosing a higher number of channels did not produce remarkable improvement.
A pooling of stride two was chosen considering two facts - (i) greater number of strides leads to insufficient number of features available in 50Hz sample frequency data for two pairs of convolution-pooling blocks, and (ii) higher number of strides may lead to information loss in time series data.

\subsubsection{Fully covolutional network parameters}
FCN employs multiple convolution blocks (one conv block = convolution + batch-normalisation + activation) and the minimum configuration is to use a single such block.
A model of three convolution blocks was found sufficient (i.e. no remarkable performance improvement with more convolution blocks), with the number of convolution output channels 128, 128, and 2.
The number of output channels in the last convolution layer was set to two, due to the fact that the global-average-pooling (GAP) layers reduces each channel features to a single scalar, and in this way, two scalar outputs of the GAP layer goes into the final softmax layer for the binary clasification.

\subsubsection{Dense network parameters}
DenseNet consists of a sequence of dense block and transition layer and the minimum configuration was found optimum for this study which consists of a single dense block of five convolution layers followed by a transition layer.
Each convolution layer, in the dense-block, produces 32 feature-maps (number of convolution output channels), which is concatenated, in the depth direction, to all the previous convolutions feature-maps.
For example, the second convolution layers inputs 32 channel data (output of 1st convolution layer is 32), but the third convolution layer inputs 64 channel data (i.e. $32*2=64$ for two previous convolution layers) and produces 32 channel output, and so on.
The number of convolution output channels was set to 32 as a base value and further optimisation could be possible, however, no significant improvement was encountered using greater number of channels or greater number of dense block and transition layers.
The transition layer performs a convolution using 1x1 kernel to capture across-the-channel features and produces the same 32 channel output feature-maps, which then goes through an average pooling layer for sub-sampling.
The number of output neurons in the hidden dense-layer was not fixed to a single value, rather, it was set dynamically as the half of the number of flattened features from the previous layer.
This decision was made because with different number of flattened number of features (due to different sample frequency data), the same number of neurons (in the hidden dense layer) may not be proportionate across different input resolution data (due to different sampling frequency).
For example, the number of neurons in the hidden dense layer of 200, may not be suitable for the flattened number of feature of 30 (i.e. for 50Hz data), but may seem fair enough for flattened number of feature of 400 (i.e., for 1000Hz data).

\subsubsection{Discriminative filter-bank learning model parameters}
The DFL model architecture, at the end of the fourth convolution layer in the initial convolution block, outputs 64 channel feature-maps, which is then transformed into 16 channels in the G-stream convolution, and into the same 64 channels (but using 1x1 convolution) in the P-stream convolution.
After passing through a global-max-pooling, a cross-channel-pooling (in the side branch, right hand side in Figure \ref{fig:arch_four}-d) was performed by dividing 64 channels into two groups (due to binary classification) of 32 channels each.
The number of channels in each group may be optimised to be as low as 16 (or 8), or as high as 64 (or 128), however, 32 channels produced a reasonable performance with moderate network complexity (more channels lead to more trainable parameters) and no significant performance improvement was observed using the mentioned variations.

\subsubsection{Common CNN hyper-parameters}
The size of the convolution kernel was set for the initial convolution layer, following a similar strategy to \cite{guo_inter-patient_2019}, equivalent to 44 ms and this remains unchanged for all the convolution layers throughout all the CNN models.
The average duration of the QRS complex is 60 ms, thus, 44 ms equivalent samples, as a convolution kernel, should be small enough to capture most ECG features.
The convolution kernel size for 50Hz, 100Hz, 250Hz, 500Hz, 1000Hz and 2000Hz sampled data, thus, consist of 3, 5, 11, 23, 45 and 89 number of samples.

The number of epoches was fixed to ten for all the models at each sample frequency.
It can be argued that different models may converge at different rates during the training process, which may require variable number of epoches.
However, in this study, CNN models were treated equally to better compare their capabilities against a common set of parameters.

\subsection{Model Testing}
Both intra and inter-database testing were carried out in this study, following approaches similar to the study \cite{habib_impact_2019}, where the intra-database tests reveal the subject-wise generalisabilty within the same database, and the inter-database tests explore a model's generalisation power across unknown validation databases with, possibly, different recording environment and patient-level data variance.
The segmentation process yields more negative segments from each recording.
To maintain the class balance, the policy of undersampling the majority class (negative segments) was adopted (negative labeled segments were randomly selected and removed from the training dataset).
All the simulations were carried out in a multi-GPU server (with GeForce GTX 1080 Ti) sequentially in order to avoid any impact that may arise due to concurrency.
Ideally, most of these simulations may be possible to run in a computer with limited resources, however, due to the loading of training and validation datasets (single training and two validation datasets per model per sampling frequency simulation) into memory may limit the execution of a comparatively complex model (DFL and DenseNet) simulation.

\subsubsection{Intra-database test}
Intra-database testing was performed using subject-wise five-fold cross validation.
Being subject-specific, the five-fold becomes unbiased, because the fraction of a validation record does not involve in the model training.
A particular model performs such an intra-database testing for each of the six sampling frequency version of a dataset.
For example, the classical convolution-pooling model performs five-fold cross validation on MIT-BIH arrhythmia dataset for six different target frequencies (MIT-BIH 50Hz, 100Hz, 250Hz, 500Hz, 1000Hz, and 2000Hz).
The best model was the one that persisted from the epoches for each fold of the five-fold operation (yielding five such best models) for later use with the cross-database validation, which is likely to enhance the cross-database validation scores.

\subsubsection{Inter-database test}
Inter-database or cross-database testing, on the other hand, was used to analyse the impact of the model's generalisation capability on six different sample frequency variations of the datasets.
Three datasets (MIT-BIH, INCART and QT) were used in this study where a model, which was trained using one particular dataset (say, MIT-BIH), gets a chance to be validated against the other two datasets (INCART and QT).
Since six different sampling frequency versions of a dataset exists, so, in this study, a particular model (say, classical convolution-pooling) trained on six different sample frequency variations of a dataset (say, MIT-BIH 50Hz, 100Hz, 250Hz, 500Hz, 1000Hz, and 2000Hz) and then validated on the six corresponding sample frequency versions of other two datasets (INCART and QT 50Hz, 100Hz, 250Hz, 500Hz, 1000Hz, and 2000Hz).
For a particular training database, sampled at one of the six frequencies, five different persisted best trained models (achieved during the five-fold intra-database validation process) were used to validate the other two validation datasets and finaly, five such validation accuracy values (per validation database) were averaged to report final validation score for that validation database.
In this way, it was expected to explore the relationship between the model's generalisation capacity and the sampling frequency of the signal for the QRS detection task.

%% file: sections/ecgdata.tex
From the PhysioNet physiological and clinical data collection \cite{goldberger_physiobank_2000}, three publicly available datasets (MIT-BIH arrhythmia, INCART, and QT) were used in this study.
The MIT-BIH arrhythmia database is sampled at 360Hz and contains 48 half-hour two channel recordings.
The INCART is a twelve-lead 75 recordings of 30 minutes length which is sampled at 257Hz, whereas, the QT consists of two-channel fifteen minutes recording of 105 subjects (usable only 82 recordings, as the annotation files available) sampled at 250Hz.
This study used single lead data from the recordings of each dataset, where channel-1 data used from INCART and QT datasets, and for MIT-BIH, either modified-lead II or V1 lead data was used.

%% file: sections/results.tex
\subsection{Effect of sampling frequency on individual model generalisability}
This study explores a CNN-based model's generalisation capability using intra and inter-database validation.
These validation results are presented under two views - (i) individual CNN-based model's generalisability on different databases, and (ii) CNN-based models performance summarised across databases.
The individual model's generalisability is described below, whereas, the models' summarised performances are presented in the following section.

\subsubsection{Intra-database performance}

\begin{figure}[h]
  \centering
  \begin{subfigure}[b]{0.4\textwidth}
    \includegraphics[width=\textwidth]{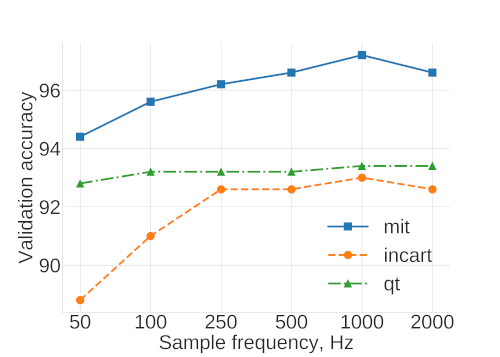}
    \caption{}
    \label{fig:convpool_intra}
  \end{subfigure}
  \begin{subfigure}[b]{0.4\textwidth}
    \includegraphics[width=\textwidth]{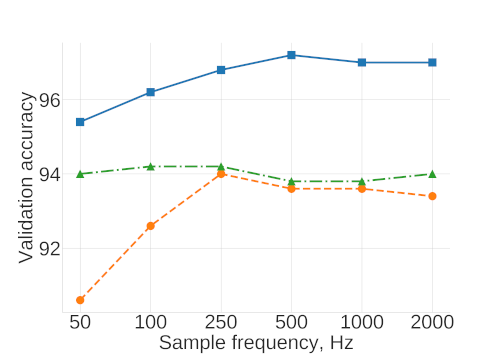}
    \caption{}
    \label{fig:fcn_intra}
  \end{subfigure}
  \begin{subfigure}[b]{0.4\textwidth}
    \includegraphics[width=\textwidth]{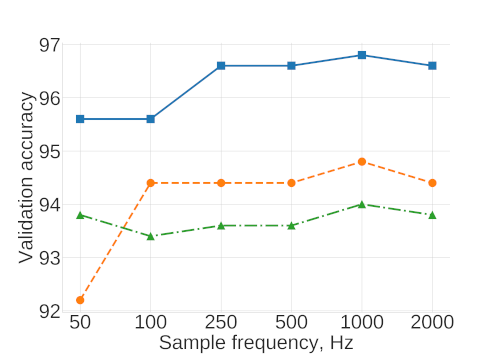}
    \caption{}
    \label{fig:dfl_intra}
  \end{subfigure}
  \begin{subfigure}[b]{0.4\textwidth}
    \includegraphics[width=\textwidth]{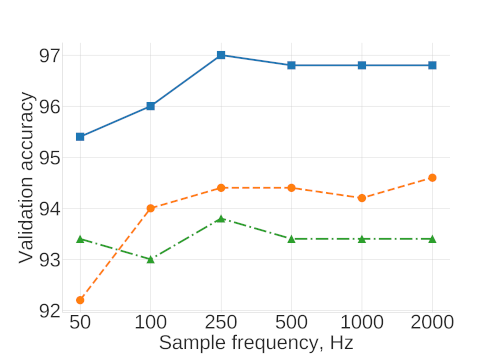}
    \caption{}
    \label{fig:dense_intra}
  \end{subfigure}
  \caption{Frequency-wise intra-database test accuracy of different CNN models, namely, (a) classical convolution-pooling, (b) fully convolutional network (FCN), (c) discriminative filter-bank learning model (DFL), (d) dense network (DenseNet) for the QRS detection task. Subject-wise five-fold validation was performed on three ECG datasets (MIT-BIH, INCART, and QT).}
  \label{fig:intra_dbwise}
\end{figure}

The results of the five-fold intra-database tests of four different CNN models operating on three datasets are represented in the Figure \ref{fig:intra_dbwise}.
MIT-BIH shows it's superiority over the other two datasets in all the CNN models (classical convolution-pooling, FCN, DFL, and DenseNet), whereas, QT is the second best for two CNN models (classical convolution-pooling and FCN), and INCART is so for the other two mdoels (DFL and DenseNet at 100Hz onwards).
250Hz is the turning point for two models (classical convolution-pooling and FCN) where INCART reaches it's peak (Figure \ref{fig:intra_dbwise}-a, b), whereas, 100Hz is the turning point for other two models for INCART reaching it's peak and at the same time, surpasses QT(Figure \ref{fig:intra_dbwise}-c, d).

\subsubsection{Inter-database performance}

\begin{figure}[h]
  \centering
  % \begin{subfigure}[b]{0.4\textwidth}
  %   \includegraphics[width=\textwidth]{convpool_inter}
  %   \caption{}
  %   \label{fig:convpool_inter}
  % \end{subfigure}
  % \begin{subfigure}[b]{0.4\textwidth}
  %   \includegraphics[width=\textwidth]{fcn_inter}
  %   \caption{}
  %   \label{fig:fcn_inter}
  % \end{subfigure}
  % \begin{subfigure}[b]{0.4\textwidth}
  %   \includegraphics[width=\textwidth]{dfl_inter}
  %   \caption{}
  %   \label{fig:dfl_inter}
  % \end{subfigure}
  % \begin{subfigure}[b]{0.4\textwidth}
  %   \includegraphics[width=\textwidth]{dense_inter}
  %   \caption{}
  %   \label{fig:dense_inter}
  % \end{subfigure}
  \begin{subfigure}[b]{0.9\textwidth}
    \includegraphics[width=\textwidth]{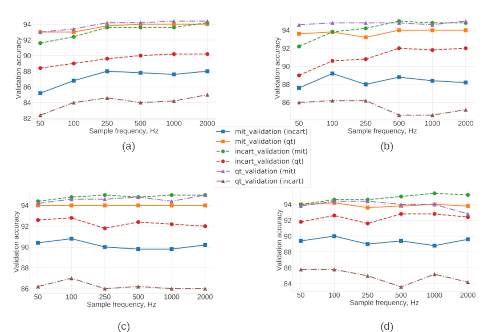}
    % \caption{}
    \label{fig:dense_inter}
  \end{subfigure}
  \caption{Frequency-wise cross-database validation accuracy of different CNN models, namely, (a) classical convolution-pooling, (b) fully convolutional network (FCN), (c) discriminative filter-bank learning model (DFL), (d) dense network (DenseNet). Three different ECG datasets (MIT-BIH, INCART, and QT) were used in testing, where a CNN model that is trained with a dataset, later used to validate against other two datasets. Pair of lines with similar line style (solid, dashed or dash-dotted) represents pair of validation datasets for a training dataset that can be found in the legend accordingly. For example, a solid line indicates a model trained with MIT-BIH dataset and the solid-blue, and solid-yellow lines indicate two validation datasets (INCART, and QT) accordingly. The best model was retrieved per fold (from the five-fold intra-database testing) and a cross-database validation was performed, yielding five cross-database validation scores which were then averaged.}
  \label{fig:inter_dbwise}
\end{figure}

The Figure \ref{fig:inter_dbwise} shows frequency-wise inter-database validation accuracy of each CNN model (average of five cross-database validation scores, received during each fold of the five-fold intra-database validation, per frequency).
The superiority of the INCART as a training dataset is visible which yields the maximum validation score for the MIT-BIH validation dataset for all the models, except for the classical convolution-pooling model where the QT as a training dataset achieves the maximum validation score for the MIT-BIH.
At 50Hz, the validation scores are at the minimum levels, and the validation performance improvement beyond the 100Hz sampling frequency (sometimes 250Hz, for example, in classical convolution-pooling model the mit-validation-incart line) is marginal.

\subsection{Effect of sampling frequency on performance of CNN models}
The performance of four CNN-based models were grouped together per sample frequency in order to observe comparative behavior clearly.
In addition, database-wise variations were further summarised into a single value (i.e. median value) for the CNN-based models for different sample frequencies.
These representations were performed for both intra and inter-database validations as below.

\subsubsection{Intra-database performance}

\begin{figure}[h]
    \centering
    \begin{subfigure}[b]{0.4\textwidth}
      \includegraphics[width=\textwidth]{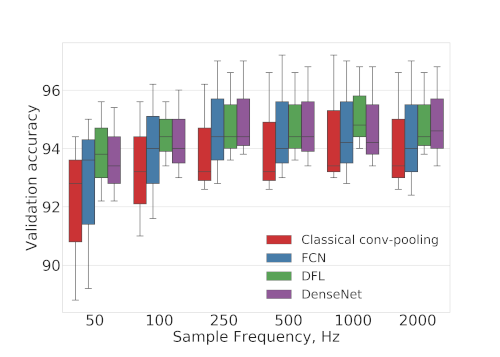}
      \caption{}
    \end{subfigure}
    \begin{subfigure}[b]{0.4\textwidth}
      \includegraphics[width=\textwidth]{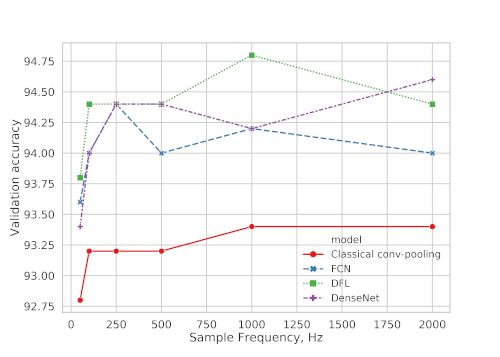}
      \caption{}
    \end{subfigure}
    \caption{(a) Median interquartile-range box-plot represents sampling frequency-wise five-fold validation accuracy grouped by four different convolutional network (CNN) based-model architectures (classical conv-pooling, fully convolutional network, discriminative filter-bank learning, and dense-network) for QRS detection task, and (b) represents the median accuracy lines summarizing median effects (of three intra-database validation scores across three datasets) at each frequency. Record-wise five-fold validation accuracy values were taken from three datasets (MIT-BIH arrhythmia, INCART, and QT) per CNN model architecture per sampling frequency. For example, the red box at 50Hz represents quartiles of three K-fold validation accuracies (of MIT-BIH, INCART and QT datasets accordingly) for classical conv-pooling CNN architecture, where, the min, median and max values are 88.8\%, 92.8\%, and 94.4\% respectively.}
    \label{fig:performance_intra_db}
\end{figure}

The results of intra-database testing, represented as the median interquartile-range box-plots in the Figure \ref{fig:performance_intra_db}-a, are the five-fold average values of the three ECG datasets (MIT-BIH, INCART, and QT) which are grouped for each sampling frequency for the four different network architectures.
There is an increasing trend of the accuracies, represented as the upper quartiles, from the lower to the higher sample frequencies for all the models, except for the case of 2000Hz, where it is slightly decreasing for two models (classical conv-pooling and DFL).
The increase of accuracy from 50Hz to 100Hz is steeper compared to the increase from 100Hz to 250Hz, and the increase, for the higher frequencies (i.e. 500, 1000 and 2000Hz), is marginal (excepting the above mentioned two cases).
In the classical conv-pooling model, for example, the peak accuracy increased 1.2\% from 50Hz to 100Hz and half (i.e. 0.6\%) from 100Hz to 250Hz, whereas successive increase rates are either marginal (i.e. 0.4\% from 250-500Hz), equal (i.e. 0.6\% from 500-1000Hz) or negative (i.e. -0.6\% from 1000-2000Hz).
The negative increase rate of -0.6\% from 1000Hz to 2000Hz in the classical conv-pooling model follows the similar decreasing pattern for the DFL model (i.e. -0.2\%), however, for the FCN and DenseNet, the increasing rate is 0\% and 0.2\% respectively.
The interquartile range (represented as solid boxes in the figure) is smaller at 100Hz (for DenseNet and DFL) and at 250Hz (for alternate conv-pooling).

The median accuracy line (Figure \ref{fig:performance_intra_db}-b) summarises (hiding database wise variations) the intra-database median behavioral response of each model on different sample frequencies.
Classical convolution-pooling model rises to the peak at two steps, first at 100Hz and then at 1000Hz, with an increase of less than 0.25\%.
The DFL also follow the same two-step rising pattern (at 100hz and 1000Hz) with an increase of less than 0.30\%, as well as, the DenseNet that touches the peak at 250Hz and then at 2000Hz.
The Fully Convolutional network shows its superiority at 250Hz, and the DenseNet, on the otherhand,

\subsubsection{Inter-database performance}

\begin{figure}[h]
    \centering
    \begin{subfigure}[b]{0.4\textwidth}
      \includegraphics[width=\textwidth]{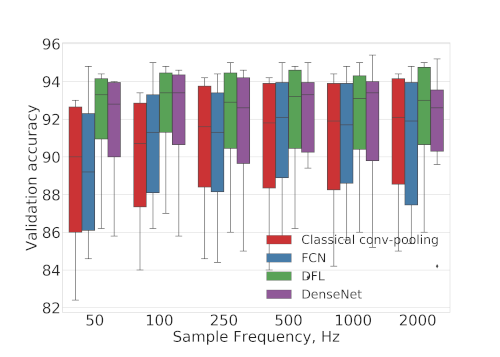}
      \caption{}
    \end{subfigure}
    \begin{subfigure}[b]{0.4\textwidth}
      \includegraphics[width=\textwidth]{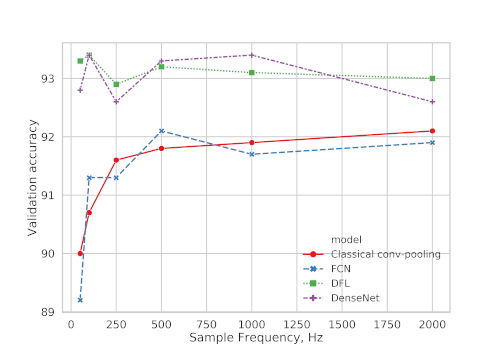}
      \caption{}
    \end{subfigure}
    \caption{(a) Median interquartile-range box-plot represents sampling frequency-wise cross database validation accuracy grouped by four different CNN model architectures (i.e. classical conv-pooling, fully convolutional network, discriminative filter-bank learning, and dense-network) for QRS detection task, and (b) represents the median accuracy lines summarizing the median effects at each frequency (of six cross-database validation scores across three datasets). Cross database validation was obtained for three different databases (i.e. MIT-BIH arrhythmia, INCART, and QT), where, for each training database, other two databases are for validation. For each sampling frequency and CNN architecture, six such cross validation accuracies can be obtained. For example, the red box at 50Hz represents the quartiles of six validation accuracy values (i.e. two validation accuracy for each of the MIT-BIH, INCART, and QT respectively) for the classical conv-pooling architecture, where, the min, median and max values represent 82.4\%, 90.0\% and 93.0\% respectively. The (b) part calculates the median of these six accuracy values per sample frequency.}
    \label{fig:performance_inter_db}
\end{figure}

The result, represented as the median interquartile-range box-plots in the Figure \ref{fig:performance_inter_db}-a, summarizes the cross database accuracy of the models for different sampling frequencies.
The upper quartiles of the interquartile range is the lowest at 50Hz for all the four models.
The upper quartiles are at the highest levels at 100hz for DenseNet, at 500Hz for FCN, and at 2000Hz for the rest two models (classical conv-pooling and DFL).
% The DenseNet at 100Hz shows the lowest deviation from the lower and upper quartiles.
The interquartile range is shorter at 100Hz (classical convolution-pooling, FCN, and DFL), and at 2000Hz (DenseNet).

The median accuracy line (Figure \ref{fig:performance_inter_db}-b) summarises (hiding database wise variations) the cross-database median behavioral response of each CNN model on different sample frequencies.
At 100Hz, the DenseNet and DFL shows the highest cross-database accuracy, although, DenseNet is very close at two other frequencies (500 and 1000Hz) and drops afterwards.
At 250Hz, the classical convolution-pooling hits the peak and then has a marginal increase with the increase in frequencies.
The FCN hits the peaks at 100Hz and 500Hz, then shows a marginal increase from 1000Hz onwards, after a decreasing period starting from the 2nd peak.

\subsection{Model costs}
Deep learning models have costs which were observed in terms of the number of trainable parameters (Table \ref{tab:tbl_model_trainable_params} and Figure \ref{fig:model_trainable_params}-a) and the training time (Figure \ref{fig:model_trainable_params}-b).

\begin{table}[h]
    \centering
    \caption{The number of trainable model parameters for different CNN models.}
    \begin{tabular}{|l|c|c|c|c|}
        \hline
        Hz & Classical Conv-Pool & FCN & DFL & DenseNet \\
        \hline
        50 & 872 & 1542 & 54440 & 64786 \\
        \hline
        100 & 1432 & 2310 & 89832 & 177874 \\
        \hline
        250 & 3362 & 4614 & 192808 & 832434 \\
        \hline
        500 & 6222 & 9222 & 298930 & 3148882 \\
        \hline
        1000 & 12132 & 17670 & 506638 & 12040994 \\
        \hline
        2000 & 23952 & 34566 & 994238 & 47105218 \\
        \hline
    \end{tabular}
    \label{tab:tbl_model_trainable_params}
\end{table}

\subsubsection{Trainable number of parameters}
\begin{figure}[h]
    \centering
    \begin{subfigure}[b]{0.4\textwidth}
      \includegraphics[width=\textwidth]{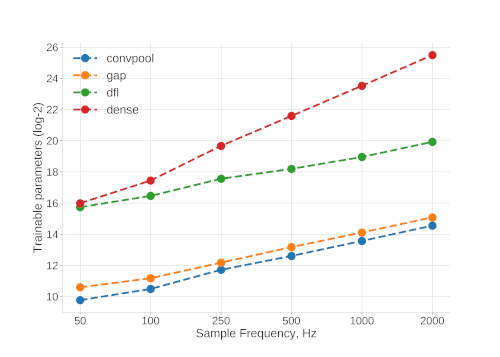}
      \caption{}
    \end{subfigure}
    \begin{subfigure}[b]{0.4\textwidth}
      \includegraphics[width=\textwidth]{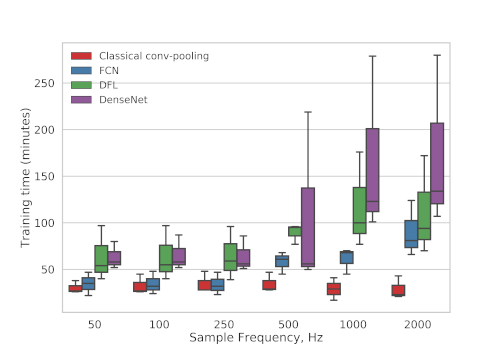}
      \caption{}
    \end{subfigure}
    \caption{(a) The number of trainable model parameters for different CNN models (i.e. Alternate conv-pooling, FCN, DFL, and DenseNet) for each sampling frequency. The number of parameters vary due to the fact that the convolution filter size grows proportionately with the increase of the sampling frequency. Due to a large variation in the number of model parameters in different models, the y-axis values are taken in $log_2$ scale. Actual parameter numbers are mentioned in Table \ref{tab:tbl_model_trainable_params}. (b) Median interquartile range box-plot represents sampling frequency-wise different CNN models training time combining three datasets (MIT-BIH, INCART, and QT). For example, at 50Hz sampling frequency, the training time of the DenseNet with three datasets are 49, 59, and 98 minutes, which are shown as the min, median and the max values in the blue box.}
    \label{fig:model_trainable_params}
\end{figure}

The Table \ref{tab:tbl_model_trainable_params} shows the number of trainable parameters of four different deep learning models operating on datasets sampled at six different frequencies and the Figure \ref{fig:model_trainable_params}-a plots these values in $log_2$ scale.
The alternate conv-pooling model has the lowest number of parameters to train at all the sample frequencies, which is relatively closely followed by FCN and distantly followed by DFL and DenseNet respectively.

\subsubsection{Training time}
% \begin{figure}[h]
%     \centering
%     \includegraphics[width=10cm]{train_time}
%     \caption{Median interquartile range box-plot represents sampling frequency-wise different CNN models training time combining three datasets (MIT-BIH, INCART, and QT). For example, at 50Hz sampling frequency, the training time of the DenseNet with three datasets are 49, 59, and 98 minutes, which are shown as the min, median and the max values in the blue box.}
%     \label{fig:train_time}
% \end{figure}

Figure \ref{fig:model_trainable_params}-b represents the training time (in minutes) of the models (for three different datasets, summarised as a single colored box, representing an interquartile range) for different operating frequencies.
As shown in the figure, there is a slowly incresing trend of the quartiles for all the CNN model architectures from 50Hz to 250Hz and a rapid increasing trend from 500Hz onwards (excepting the classical conv-pooling model which remains almost at the same level).
There is an exponential increase of the training time from 250Hz to 500Hz to 1000Hz for the DFL and DenseNet.
The FCN has a sharp increase in the training time from 250Hz to 500Hz and from 1000Hz to 2000Hz, whereas, it remains almost at the same level in the lower frequencies (50Hz, 100Hz, 250Hz, and 500Hz).
The classical conv-pooling model shows almost similar levels of the upper quartiles for all the sampling frequencies.

%% file: sections/discussion.tex
The findings of this study are discussed below.

\subsection{Intra-database performance}
The intra-database test results (Figure \ref{fig:intra_dbwise}, and Figure \ref{fig:performance_intra_db}) summarise the fact that different CNN models tend to gain increasing detection accuracies, but at a different rate, with an increase in the sampling frequency.
The rate of increase is steeper from 50Hz through to 250Hz, compared to the higher frequencies, with each of the four CNN models showing almost similar patterns, although the magnitude of the accuracy increase changes at different proportions.
Finer resolution input data (higher sample frequency) seems not capable of increasing the model generalisation proportionately, which could be due to the fact that by taking an increasing number of samples of the QRS complex and non-QRS regions, marginal or no information gain may be achieved (as shown in the Figure \ref{fig:ecg_data_viz}).
The lower accuracy levels at as low as 50Hz, on the other hand, indicates that information, at some extend, has been lost due to the lower sampling rate that keeps the CNN models from achieving the highest accuracy levels at these lower frequencies.
One particular case is the QT database, for which, the accuracy at 50Hz is competitive to the other frequencies, which indicates that very little or no information gain is taking place with an increase in the sampling frequency, for that case.

\subsection{Cross-database performance}
The cross-database validation datasets, for each training dataset, of all the CNN models, maintains one kind of ordering (as shown in the Figure \ref{fig:inter_dbwise}, qt-validation INCART line having the lowest and qt-validation MIT-BIH line with the highest accuracy values).
Each model may have superiority over the others which is realised by observing the shifting of the validation accuracy lines at different magnitude across different frequencies, keeping this ordering unchanged.
The consistent relative orderings of six different cross-database validation lines across different CNN models may be due to the fact that it is the very nature of the data itself which limits the maximum generalisability of a CNN model (irrespective of CNN architectures), keeping all other factors unchanged.
Possibly, the uniqueness of the validation datasets (i.e., subject variance and signal noise) were not present in the training dataset to allow a CNN model to exceed this generalisability limitation, and this similar observation was discussed in a study \cite{habib_impact_2019}, where it was concluded that the generalisation capability does not increase by simply adding more training samples, rather a diverse range of recordings should be included incorporating as much relevant patterns as possible.
Another sutdy \cite{guo_inter-patient_2019} reiterates this fact stating that inclusion of as much heterogeneities as possible into the training data, possibly may increase a model's generalisation ability.

\subsection{Performance of CNN models}
Interestingly, the cross-database validation accuracy at, as low as, 100Hz is the maximum for two CNN models (DFL and DenseNet), and at 250Hz and 500Hz for the classical convolution-pooling and FCN respectively (Figure \ref{fig:performance_inter_db}-b).
The variation of the cross-database validation accuracies (Figure \ref{fig:performance_inter_db}-a) summarizes that the lowest accuracy variation (low interquartile range) is observed at as low as 50Hz and 100Hz (for DFL model), although the upper quartile is the highest at higher frequencies (i.e. at 2000Hz with DenseNet).
Also, this variation, at worst case, is less than 2\% (for FCN, in Figure \ref{fig:performance_inter_db}-b).
CNN models, thus, may be capable of offering stable generalisation at lower frequency signals (i.e. 50, 100, or 250Hz), whereas, the data with the higher frequency introduces more variation.
This phenomena may be justified by the fact that the lower frequency sampled data represents the detection artifacts (i.e. a QRS complex) with the minimum input resolution, yielding a lesser noise attenuation, whereas, the higher frequency data may introduce a significant proportion of noise (keeping the basic artifact morphology similar) that might lead to inconsistent generalisation for some of the CNN models.
Another instance supports this phenomena, for example, the only CNN model that shows consistent intra-database performance improvement in higher frequencies (i.e. DenseNet at 1000Hz onwards, in Figure \ref{fig:performance_intra_db}-b), shows an opposite behaviour in the cross-database validation, where it consistently decreases with an increase in the sample frequency (in Figure \ref{fig:performance_inter_db}-b).
Figure \ref{fig:ecg_data_viz} probably justifies this phenomena which shows that the higher frequency introduces much fluctuation to the data as compared to the lower sample frequencies.
The figure also justifies the fact that the lower frequency may significantly distort the morphology of the signal artifacts (the third beat as shown in the Figure \ref{fig:ecg_data_viz}-a) which may lead to poor detection performance and thus poor generalisation capability of CNN models.

% The inter-database testing follows almost the similar performance pattern of the intra-database testing, where the accuracy levels (in the Figure \ref{fig:inter_dbwise}, as well as, represented as the upper quartile of the interquartile range in Figure \ref{fig:performance_inter_db}) increase with an increase in the sample frequency from 50Hz till 500Hz, however, some models (DFL and DenseNet) get their accuracies varied at higher frequencies.

% Both the intra and inter-database test results suggest that the CNN model's performance is positively related to the sample frequency (of the data), however, this is not linear, since lower frequencies (100Hz and 250Hz) has higher slope compared to the increase line of the higher frequencies (i.e. 500Hz onwards).
% The difference of accuracy levels from 100Hz to the rest is less than 1\%, which favors choosing comparatively lower frequencies (100Hz or 250Hz) for the QRS detection task without losing much accuracies.

\subsection{Model cost and performance}
With the increase of sampling frequency, CNN models input resolution and convolution filter size changes, yielding an exponentially increasing number of trainable model parameters (model complexity, in Figure \ref{fig:model_trainable_params}-a).
Training a CNN model with an increasing number of parameters may require an increasing number of training samples and computation resources.
It is also clear that (as shown in the Figure \ref{fig:model_trainable_params}-b) the CNN models with higher number of trainable parameters (yielding from higher sample frequency data) require comparatively higher training time.
The variation of training time around 100Hz and 250Hz is negligible, however, for some models (DFL and DenseNet), it grows exponentially at 500Hz and 1000Hz.
The negligible increase of the training time in lower sample frequency simulation may be justified by the fact that for a dedicated GPU of 10 giga-byte memory, the simulations of 50, 100 and 250Hz data failed to fully occupy the allocated computation resource compared to higher frequency simulations (500Hz onwards).
The computed training time would have been different for the lower frequency data simulations (as well as the higher counterpart), if they were executed in a computer without a GPU.

\subsection{Implications}
Different application settings may require different detection accuracy levels and computation resources.
For example, the QRS detection in wearables may prefer a light-weight solution where sacrificing 1\% accuracy may not have a severe impact and can go for data sampled at as low as 50Hz, compared to the deployment in the clinical settings (e.g., in an intensive care unit), where a higher detection rate may be critical, thus, adoption of ECG data sampled at higher frequencies may be preferrable, which is trained with a better generalising model (e.g. DenseNet or other).

% Both intra and inter-database testing suggests that frequencies as low as 50Hz, 100Hz or 250Hz yields an acceptable level of accuracy (while 50Hz shows comparatively the lowest performance at a margin around 2\%) and higher frequency levels contributes accuracy gain marginally (introducing more variations), however, such marginal increase may require a CNN model with a higher number of trainable parameters and higher training time.

Finally, the findings of this CNN-based QRS detection study can be summarised as below:
\begin{itemize}
\item Sample requency as low as 50Hz, 100Hz or 250Hz yields an acceptable level of accuracy (while 50Hz shows comparatively the lowest performance at a margin around 2\%)
\item Higher sample frequencies contribute marginally to the accuracy gain (introducing more variations), however, such marginal increase may require a CNN model with a higher number of trainable parameters and higher training time.
\item Some CNN model architectures show greater generalisability (i.e. DenseNet or DFL) compared to others (i.e. classical convolution-pooling or FCN).
\end{itemize}

%% file: sections/conclusion.tex
Detection of the QRS complex is necessary for automated ECG signal analysis and with the increasing usage of wearables in the home-based personal health monitoring, a careful selection of the data sampling frequency may be necessary in order to obtain a simpler deep learning based model.
In this experimental research, the effects of data sampling frequency were investigated using three datasets (MIT-BIH, INCART, and QT) resampled at six different sample frequencies (50, 100, 250, 500, 1000, and 2000Hz) following both intra and inter-database testing methods.
Four different CNN-based models (classical convolution-pooling, fully convolutional network, discriminative filter-bank learning, and dense network) were used to justify the findings.
The results show that the model accuracy has a positive relationship with the sample frequencies across all the models at a varying magnitude.
Data signal sampled at as low as 50Hz seems capable of representing ECG signal morphologies (the QRS complex), losing information (as well as detection accuracy) at a certain level, keeping the CNN model simple, as compared to the higher frequency counterparts, where increased signal resolution may not add much information proportionately.
In addition, the higher frequency signals may have a chance of getting distorted due to the resampling method (i.e. interpolation).
Lower resolution signals also show greater generalisability of the CNN model (i.e. at 100Hz or 250Hz), compared to the higher resolution signals (i.e. 2000Hz).
The QRS detection accuracy variation is in the range of 2-3\% from 100Hz to the higher frequency levels, however, the increase of the model complexity is much higher in the higher frequencies.
To carefully design a training dataset which consists of enough heterogeneities for a suitable CNN-based model to have better generalisability could be one of the future tasks.